\documentclass[pdftex,twocolumn,epjc3]{svjour3}          

\RequirePackage[T1]{fontenc}

\smartqed  

\RequirePackage{graphicx}
\RequirePackage{mathptmx}      
\RequirePackage{flushend}
\RequirePackage[numbers,sort&compress]{natbib}
\RequirePackage[colorlinks,citecolor=blue,urlcolor=blue,linkcolor=blue]{hyperref}
\usepackage{amsmath, amssymb}
\usepackage{slashed}
\usepackage{graphicx}
\usepackage{amsmath}
\usepackage{amssymb}
\usepackage{bbold}
\usepackage{wrapfig}
\usepackage{color}
\usepackage{hyperref}
\DeclareMathOperator{\Tr}{Tr}

\journalname{Eur. Phys. J. C}

\begin{document}

\title{Multicomponent superfluidity in two-color QCD at finite density at next-to-leading order}


\author{Prabal Adhikari\thanksref{e1,addr1}
        \and
        Huy Nguyen\thanksref{e2,addr2} 
}

\thankstext{e1}{e-mail: prabal.adhikari@wellesley.edu}
\thankstext{e2}{e-mail: nguyen28@stolaf.edu}

\institute{Wellesley College, Department of Physics, 106 Central Street, Wellesley, MA 02481, USA\label{addr1}
          \and
          St. Olaf College, 1520 St. Olaf Av., Northfield, MN 55057, USA\label{addr2}
}

\date{Received: date / Accepted: date}

\maketitle

\begin{abstract}
In this paper, we study two-color, two-flavor QCD using chiral perturbation theory at next-to-leading order when the diquark chemical potential ($\mu_{B}$) is equal to the isospin chemical potential ($\mu_{I}$). For chemical potentials larger than the physical pion mass, the system is in a multicomponent superfluid phase with both pions and diquarks. We construct the one-loop effective potential using $\chi$PT in the presence of an external multicomponent superfluid source and use the effective potential to calculate the chiral condensate, the multicomponent superfluid condensate and the (multicomponent) superfluid density. We also find the critical chemical potential and the order of the phase transition from the normal phase to the multicomponent condensed phase at next-to-leading order. The phase transition remains second order (as at tree-level) and the critical chemical potential is equal to the one-loop renormalized diquark (or pion) mass.
\end{abstract}

\section{Introduction}
\label{Introduction}
Quantum Chromodynamics (QCD) has a rich phase structure but is notoriously difficult to solve due to its non-pert-urbative nature and the presence of the fermion sign problem in lattice QCD~\cite{Barbour:1998he} at finite baryon density (relevant for nuclear matter and neutron stars) and at finite isospin densities with external magnetic fields. Due to these challenges, it is often insightful to study variants of QCD. For instance, two-color QCD at finite baryon density (or finite isospin density) does not suffer from the sign problem and can be studied using lattice QCD, where it possesses a diquark condensate both at low and high baryon densities~\cite{Iida:2019rah,Braguta:2016cpw}. This is different from three-color QCD, which supports nuclear matter at low densities and the color-flavor-locked (CFL) phase at asymptotically large densities~\cite{Alford:1998mk}. Large $N_{c}$ QCD, on the other hand, possesses a phase structure qualitatively different from that of real QCD -- at finite baryon densities the deconfinement transition temperature is independent of the baryon chemical potential~\cite{Toublan:2005rq}. While quark loops are suppressed in the 't Hooft large $N_{c}$ limit, in real QCD, quark loops screen the interactions of valence quarks leading to the decrease of the deconfinement temperature with increasing baryon density. Furthermore, large $N_{c}$ QCD with heavy quarks allows for the possibility of saturated nuclear matter, which forms for qualitatively different reasons than real QCD: nuclear matter saturates due to the competition between Pauli repulsion (which is leading order in $N_{c}$) and the sub-leading (attractive) glueball interactions resulting in a nuclear saturation density that is non-analytic in $N_{c}$~\cite{Adhikari:2013dfa}.

Unlike QCD at finite baryon chemical potentials, it is possible to study QCD at finite isospin density both at low densities (compared to the typical hadronic scale $\Lambda_{\rm hadron}\sim 4\pi f_{\pi}$) using chiral perturbation theory ($\chi$PT)~\footnote{For a review, see \cite{scherer2011primer, Pich:2018ltt} and for seminal literature, see Refs.~ \cite{Gasser:1979hf,Gasser:1982ap,Gasser:1983kx,Gasser:1983yg,Gasser:1984gg,Gasser:1984ux,Gasser:1986pc,Gasser:1986vb,Gasser:1987ah}} and (in principle) at all isospin densities using lattice QCD~\cite{Brandt:2017oyy,Kogut:2002se,Sinclair:2003rm}. The argument is that at finite isospin density (in the absence of a magnetic field) the complex phase that arises due to the chemical potential of up quarks is cancelled exactly by the opposite phase carried by the down quarks due to its chemical potential. Lattice QCD, similar to finite isospin chiral perturbation theory ($\chi$PT), predicts the existence of a phase transition to a pion superfluid, which at high densities becomes a Bardeen-Cooper-Schrieffer (BCS) condensate though there is no change in the symmetries of the order parameter and hence no real phase transition, only a crossover transition to a ``weaker" condensate. However, at finite magnetic fields, the charge asymmetry of the quarks leads to a non-zero complex phase and hence the re-emergence of the sign problem~\cite{Endrodi:2014lja}. 

While pion superfluidity in QCD~\cite{Son:2000xc,Son:2000by,Carignano:2016lxe} is an example of a relativistic superfluid similar in principle to the superfluidity in the Abelian Higgs Model, the first experimental observation of superfluidity was in the context of Helium-4, a non-relativistic superfluid (for a review, see Ref.~\cite{Mannarelli:2019hgn}). Landau provided the first explanation of superfluidity using the notion of quasiparticles. Within modern field theory, superfluidity (in both relativistic and non-relativistic systems) arises due to the breaking of a global $U(1)$ symmetry. Since pions are electromagnetically charged, in the presence of an external magnetic field they exhibit superconductivity and form magnetic vortices as has been shown using finite isospin $\chi$PT~\cite{Adhikari:2015wva,Adhikari:2018fwm}. Superconductivity was first explained through Ginzburg-Landau (GL) theory~\cite{Ginzburg:1950sr} and later further clarified through BCS theory and the Cooper pairing mechanism whereby two electrons couple through (attractive) phonon interactions~\cite{Bardeen:1957kj,Bardeen:1957mv} with Abrikosov recognizing the possibility of magnetic vortices and type-II superconductivity~\cite{Abrikosov:1956sx}. A modern field theoretic understanding via the Abelian Higgs model suggests the breaking of a local $U(1)$ symmetry as the mechanism that gives rise to both type-I and type-II superconductivity~\cite{Harrington:1975mv}. 

The phase diagram of QCD possesses more exotic possibilities than single species superfluids and superconductors. For instance, $\rho$ mesons are expected to condense at large magnetic fields~\cite{chernodub2012electromagnetically,Chernodub:2012tf,Chernodub:2014rya}, with magnetic vortices involving multiple species, both electromagnetically charged and neutral condensates. Similarly, it is expected that neutrons form a superfluid in neutron stars with protons forming a superconducting state~\cite{Schmitt:2010pn,Haber:2017kth}. Outside of QCD, in condensed matter systems, the existence of exotic vortex phases is known in superfluid He-3~\cite{Achucarro:1999it} and recently multicomponent mesonic systems were studied using $\chi$PT in Ref.~\cite{Lepori:2019vec}. 

The focus of this paper is two-color, two-flavor QCD at low densities, where $\chi$PT is applicable at finite density with the isospin chemical potential equal to the diquark chemical potential. The system forms a multicomponent superfluid with a condensate that is a mixture of a pion and a diquark condensate with neither fixed separately~\cite{Adhikari:2018kzh}, an issue that was ignored in the seminal literature on two-color, two-flavor $\chi$PT. The goal of this paper is to construct the next-to-leading order  (NLO) effective potential and calculate the NLO chiral and multi-component superfluid component. A recent study of two-flavor, three-color $\chi$PT in Ref.~\cite{Adhikari:2020ufo} shows that the NLO corrections are crucial in order to correctly capture the qualitative features of the pion condensate. A similar possibility exists in three-flavor, three-color $\chi$PT at finite isospin and so-called strange chemical potentials with the formation of a condensate that is a ``mixture" of the quantum numbers of both pions and kaons, with neither fixed separately by the chemical potentials~\cite{Kogut:2001id}. In this instance, quantum corrections shift the position of the first-order line~\cite{Adhikari:2019mlf} unlike what occurs in two-color, two-flavor $\chi$PT due to the symmetry of the phase diagram with respect to the exchange of isospin and baryon chemical potentials.

The paper is organized as follows. In Sec.~\ref{review}, we begin with a brief review of two-color, two-flavor $\chi$PT including the possibility of a multicomponent superfluid. In Sec.~\ref{coset}, we discuss the parameterization of the Goldstone manifold (coset space) associated with the (spontaneous) symmetry breaking pattern in two-color QCD. In Sec.~\ref{NLOLag}, we use the coset parameterization to calculate the Lagrangian in terms of the fluctuations around the multicomponent superfluid ground state. We also determine the dispersion relation and the tree level masses in Sec.~\ref{dispersionsection}. In Sec.~\ref{1loopeffectivepotentialsection}, we use the dispersion relation and the relevant counterterm Lagrangian to determine the one-loop effective potential and in Sec.~\ref{cond} we calculate the chiral condensate, the number density and the multicomponent superfluid condensate at next-to-leading order in $\chi$PT. We conclude with a discussion in Sec.~\ref{discussion}. Finally, we list some useful dimensional regularization integrals in \ref{DR}, the NLO Lagrangian with one and two field variables in \ref{L41L42} and calculate the one-loop pion self-energy and pion decay constant in the normal vacuum in \ref{selfenergy}.
\section{Review of Two-Color Chiral Perturbation Theory}
\label{review}
\noindent
We begin with a brief discussion of two-color~\cite{Kogut:2000ek,Kogut:1999iv,Splittorff:2001fy,Kogut:2001if} $\chi$PT. The theory possesses an expanded Pauli-Gursey symmetry ($SU(2N_{f})_{\rm flavor}$) due to the pseudo-reality of the color gauge group $SU(2)_{\rm color}$~\cite{Pauli:1957,Gursey:1957}. This allows for the possibility of combining quarks with charge-conjugated antiquarks in a multiplet (Weyl spinor) that transforms under an enlarged $SU(2N_{f})$ flavor group as opposed to the $SU(N_{f})\times SU(N_{f})$ flavor group of three-color QCD. The vacuum breaks the symmetry down to $Sp(2N_{f})$ resulting in a Goldstone manifold that has $2N_{f}^{2}-N_{f}-1$ physical degrees of freedom. We will focus on the two-flavor case in this paper. For $N_{f}=2$, there are five degrees of freedom corresponding to the charged diquarks ($d_{\pm}$), the charged pions ($\pi_{\pm}$) and a neutral pion ($\pi_{0}$). 

The two-color, two-flavor $\chi$PT Lagrangian at $\mathcal{O}(p^{2})$ is
\begin{equation}
\mathcal{L}_{2}=\frac{f^2}{4}{\rm Tr}\left [ \nabla_{\mu} \Sigma (\nabla^{\mu} \Sigma)^{\dagger}\right ]+\frac{f^{2}}{4} {\rm Tr}\left [\chi^{\dagger} \Sigma+\Sigma^{\dagger}\chi\right ]\ ,
\end{equation}
where $\chi$ represents external (scalar, pseudoscalar, diquark or multicomponent superfluid) sources and  $f$ is the (bare) pion decay constant. The covariant derivatives are defined as
\noindent
\begin{equation}
\begin{split}
\label{covariantderivative}
\nabla_{\mu} \Sigma&\equiv\partial_{\mu}\Sigma-i\delta_{\mu0}\left\{\mu_{B}B+\mu_{I}I_{3},\Sigma\right\}\\ 
\nabla_{\mu} \Sigma^{\dagger}&\equiv\partial_{\mu}\Sigma^{\dagger}+i\delta_{\mu0}\left\{\mu_{B}B+\mu_{I}I_{3},\Sigma^{\dagger}\right\}\ ,
\end{split}
\end{equation}
where $\mu_{B}$ is the diquark (baryon) chemical potential and $\mu_{I}$ is the isospin chemical potential. In the analysis of this paper, we use the following source term
\begin{equation}
\begin{split}
\label{sourceterm}
\chi&=\frac{G\mathcal{M}^{\dagger}}{2f^{2}},\ \mathcal{M}=\hat{m}\hat{M}+j\hat{J},\ \\
\hat{M}&=
\begin{pmatrix}
0&\mathbb{1}\\
-\mathbb{1}&0
\end{pmatrix},\ \hat{J}=
\begin{pmatrix}
\tau_{2}&0\\
0&\tau_{2}
\end{pmatrix}\ ,
\end{split}
\end{equation}
where $\hat{m}$ is the scalar (quark mass) source or the quark mass in the isospin limit and $j$ is the source for the multicomponent of diquarks and pions. The scalar source introduces quark masses in $\chi$PT, and allows us to calculate the chiral condensate using the Gell-mann-Oakes-Renner relation (GOR) relation,
\begin{equation}
\begin{split}
\label{GOR}
2G\hat{m}=m^{2}f^{2}\ ,
\end{split}
\end{equation}
where $m$ is the bare pion mass (which is degenerate with the bare diquark mass), $f$ is the bare pion decay constant, $\hat{m}$ is the quark mass in the isospin limit and $-4G$ is the chiral condensate in the normal vacuum.
Similarly, the multicomponent superfluid source allows for the computation of the condensate associated with the multicomponent superfluid phase, which occurs (as we will discuss next) when the isospin chemical potential equals the baryon chemical potential. $\mathbb{1}$ is the $2\times 2$ identity matrix and $\tau_{i}$s are the Pauli matrices.
The baryon and isospin charge matrices in Eq.~(\ref{covariantderivative}) are defined as
\begin{equation}
\begin{split}
B&=\frac{1}{2}\begin{pmatrix}
\mathbb{1}&0\\
0&-\mathbb{1}
\end{pmatrix},\ I_{3}=\frac{1}{2}\begin{pmatrix}
\tau_{3}&0\\
0&-\tau_{3}
\end{pmatrix}\ ,
\end{split}
\end{equation}
with the elements in the diagonal referring to the isospin and baryon charges of the up, down, anti-up and anti-down quarks respectively. 
\subsection{Ground State}
\begin{figure}[h]
    \centering
    \includegraphics[width=.5\textwidth]{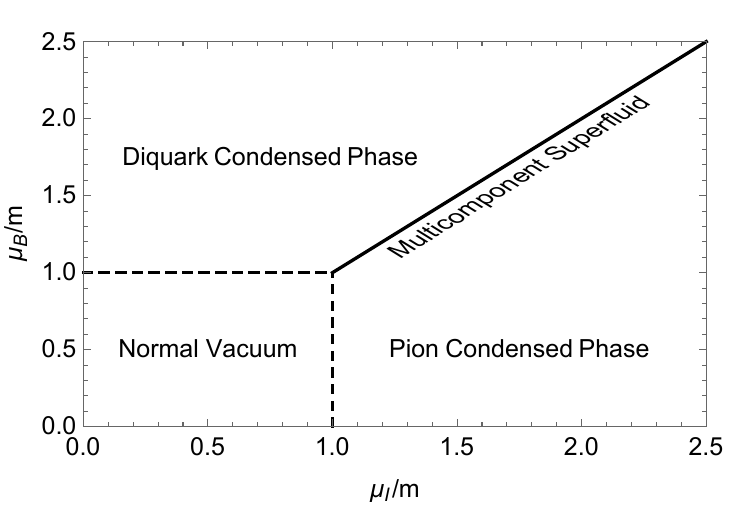}
    \caption{Phase diagram of two-color, two-flavor $\chi$PT at finite density. When $\mu_{I}<m$ and $\mu_{B}<m$, the system is in a normal phase. For $\mu_{I}>m$ and $\mu_{I}\neq \mu_{B}$, the system is in a pion condensed phase. For $\mu_{B}>m$ and $\mu_{B}\neq \mu_{I}$, the system is in a diquark condensed phase. Finally, when $\mu_{I}=\mu_{B}>m$, the system is in a multicomponent superfluid phase.}
\label{phase_diagram}
\end{figure}

\noindent
In order to parameterize the ground state, we choose the following ansatz~\cite{Adhikari:2018kzh} for the orientation of $\Sigma$,
\begin{equation}
\begin{split}
\label{gs}
\overline{\Sigma}(\alpha)&=\cos\alpha\ \Sigma_{0}+i\sin\alpha\ \Sigma_{i}\hat{\phi}_{i}\\
\Sigma_{0}&=
\begin{pmatrix}
0&-\mathbb{1}\\
\mathbb{1}&0\\
\end{pmatrix},\ 
\Sigma_{1}=
\begin{pmatrix}
-i\tau_{2}&0\\
0&-i\tau_{2}\\
\end{pmatrix}\\
\Sigma_{3}&=
\begin{pmatrix}
0&\tau_{1}\\
-\tau_{1}&0\\
\end{pmatrix},\ 
\Sigma_{2}=
\begin{pmatrix}
\tau_{2}&0\\
0&-\tau_{2}\\
\end{pmatrix}\\
\Sigma_{4}&=
\begin{pmatrix}
0&\tau_{2}\\
\tau_{2}&0\\
\end{pmatrix},\ 
\Sigma_{5}=
\begin{pmatrix}
0&\tau_{3}\\
-\tau_{3}&0\\
\end{pmatrix}\ .
\end{split}
\end{equation}
$\Sigma_{0}$ represents the orientation of the normal vacuum that breaks chiral symmetry; $\Sigma_{1}$ and $\Sigma_{2}$ represent the orientations of the (charged) diquark; $\Sigma_{3}$ and $\Sigma_{4}$ represent the orientations of the (charged) pions; and finally $\Sigma_{5}$ represents the orientation of the neutral pion. Since $\overline{\Sigma}$ is unitary, i.e. $\overline{\Sigma}^{\dagger}\overline{\Sigma}=\mathbb{1}$, the $\hat{\phi}_{i}$s obey the constraint,
\begin{equation}
\begin{split}
\label{phihat}
\sum_{i=1}^{5}\hat{\phi}_{i}\hat{\phi}_{i}=1\ .
\end{split}
\end{equation}
We use the above ansatz and the GOR relation of Eq.~(\ref{GOR}) to construct the tree level potential under the assumption of homogeneous phases and the absence of the multicomponent superfluid source ($j=0$). We get
\begin{equation}
\begin{split}
V_{\rm tree}&=\frac{f^{2}}{4}\Tr\left[\{\mu_{B}B+\mu_{I}I_{3},\Sigma\}\{\mu_{B}B+\mu_{I}I_{3},\Sigma^{\dagger}\}\right.\\
&\left.+\frac{Gm}{2f^{2}}\left(\hat{M}\Sigma+\Sigma^{\dagger}\hat{M}^{T}\right)\right]\\
&=-f^{2}\left[2m^{2}\cos\alpha\right.\\
&\left.+\sin^{2}\alpha\left\{\mu_{B}^{2}(\hat{\phi}_{1}^{2}+\hat{\phi}_{2}^{2})+\mu_{I}^{2}(\hat{\phi}_{3}^{2}+\hat{\phi}_{4}^{2})\right\}\right]\ .
\end{split}
\end{equation}
Minimizing $V_{\rm tree}$ with respect to  $\alpha$, we get 
\begin{equation}
\begin{split}
\sin\alpha\left [m^{2}-\cos\alpha\left\{\mu_{B}^{2}(\hat{\phi}_{1}^{2}+\hat{\phi}_{2}^{2})+\mu_{I}^{2}(\hat{\phi}_{3}^{2}+\hat{\phi}_{4}^{2})\right\}\right ]=0\ .
\end{split}
\end{equation}
Using this condition (and the curvature of the tree level potential), we find that there are three distinct condensed phases in addition to the normal phase. The resulting phase diagram is shown in Fig.~\ref{phase_diagram}.
\\
\begin{itemize}
\item If $\mu_{B}\le m$ and $\mu_{I}\le m$, the system is in the normal vacuum with $\overline{\Sigma}=\Sigma_{0}$, i.e. $\hat{\phi}_{i}=0$ and $\alpha=0$. \\
\item If $\mu_{B}>\mu_{I}$ and $\mu_{B}> m$, the system is in the diquark condensed phase, i.e. $\hat{\phi}_{1}^{2}+\hat{\phi}_{2}^{2}=1$ and $\hat{\phi}_{3}=\hat{\phi}_{4}=\hat{\phi}_{5}=0$.\\
\item If $\mu_{I}>\mu_{B}$ and $\mu_{I}> m$, the system is in the pion condensed phase i.e. $\hat{\phi}_{3}^{2}+\hat{\phi}_{4}^{2}=1$ and $\hat{\phi}_{1}=\hat{\phi}_{2}=\hat{\phi}_{5}=0$.\\
\item Finally, if $\mu_{B}=\mu_{I}=\mu$ and $\mu>m$, the system is in the multicomponent superfluid phase with a vacuum that is degenerate in $\hat{\phi}_{i}$, $i=1,2,3,4$, i.e. $\hat{\phi}_{1}^{2}+\hat{\phi}_{2}^{2}+\hat{\phi}_{3}^{2}+\hat{\phi}_{4}^{2}=1$ and $\hat{\phi}_{5}=0$.\\
\end{itemize}
In each of the condensed phases, $\alpha=\arccos\left (\frac{m^{2}}{\mu_{i}} \right )$, where $\mu_{i}=\mu_{I}$ in the pion condensed phase, $\mu_{i}=\mu_{B}$ in the diquark phase and $\mu_{i}=\mu_{B}=\mu_{I}\equiv\mu$ in the multicomponent superfluid phase, which is characterized by the condition $\sum_{i=1}^{4}\hat{\phi}_{i}=1$ and $\hat{\phi}_{5}=0$. There are additional degrees of freedom compared to the diquark or the pion condensed phases. This leads to an enlarged $SU(2)$ global symmetry in the Lagrangian unlike the global $U(1)$ symmetry when $\mu_{B}\neq\mu_{I}$, which is broken in the superfluid phase. For details of how the SU(2) symmetry manifests, see Ref.~\cite{Adhikari:2018kzh}.
\section{Parameterization of the Goldstone Manifold}
\label{coset}
The goal of this section is to parameterize $\Sigma$ including its fluctuations around its vacuum expectation value (vev) discussed in the previous section. This is necessary for the construction of the one-loop effective potential associated with the multicomponent superfluid phase. We begin by noting that the Lagrangian has an enlarged global symmetry $SU(2N_{f})$ which in the presence of a finite quark mass breaks down to $Sp(2N_{f})$. The Goldstone manifold is
\begin{equation}
\begin{split}
SU(2N_{f})/Sp(2N_{f})\ ,
\end{split}
\end{equation} 
and it can be parametrized in terms of the broken generators ($X_{i}$), which themselves rotate as the ground state rotates from the normal phase to one of the condensed phases. In the normal phase, $\Sigma$ including its fluctuations ($\phi_{i}$) is parameterized as~\cite{Kogut:2000ek}
\begin{equation}
\begin{split}
\label{Sigma0}
\Sigma(0)&=U_{0}\Sigma_{0}U_{0}^{T}=U^{2}_{0}\Sigma_{0}\\
U_{0}&=e^{\frac{iX_{i}\phi_{i}}{2f}}\ ,
\end{split}
\end{equation}
where $X_{i}$ are the five broken generators of $SU(2N_{f})/Sp(2N_{f})$ with $N_{f}=2$ and the explicit form of $\Sigma_{0}$ is stated in Eq.~(\ref{gs}). The relationship follows from the fact that the broken generators obey $X_{i}\Sigma_{0}=\Sigma_{0}X_{i}^{T}$~\cite{Kogut:2000ek}, which can be checked explicitly using $\Sigma_{0}$ in Eq.~(\ref{gs}) and the broken generators, $X_{i}$, in Eq.~(\ref{generators}).

Since we are interested in calculating the one-loop effective potential of the multicomponent superfluid phase in this paper, we parameterize the fluctuations around the multicomponent superfluid phase, which can be written in terms of $\Sigma_{i}$ of Eq.~(\ref{gs}) and a ``rotation'' matrix, $V_{\alpha}$, as follows
\begin{equation}
\begin{split}
\label{Sigmaalpha}
\overline{\Sigma}(\alpha)&=V_{\alpha}\Sigma_{0}V_{\alpha}^{T}\\
V_{\alpha}&=e^{-\frac{i\alpha}{2}\left (\sum_{i=1}^{4}\hat{\phi}_{i}\Sigma_{i}\Sigma_{0} \right )},
\end{split}
\end{equation}
with $\hat{\phi}_{i}$ chosen appropriately for each of the condensed phases as discussed in the previous section. The parameterization is completely general and valid in any of the condensed phases (including the normal phase for which $\alpha=0$).  Then, $\Sigma(\alpha)$ in the multicomponent superfluid phase including the fluctuations $\phi_{i}$ around $\overline{\Sigma}(\alpha)$ is
\begin{equation}
\begin{split}
\label{Sigmaalpha}
\Sigma(\alpha)&=U_{\alpha}\overline{\Sigma}(\alpha)U_{\alpha}^{T}=V_{\alpha}U_{0}^{2}\Sigma_{0}V_{\alpha}^{T}\\
U_{\alpha}&=V_{\alpha}U_{0}V_{\alpha}^{\dagger}\ ,
\end{split}
\end{equation}
which reduces to the result of Eq.~(\ref{Sigma0}) as it should in the normal vacuum.
The second equality of Eq.~(\ref{Sigmaalpha}) follows from the fact that $V_{\alpha}^{\dagger}V_{\alpha}=\mathbb{1}$ and the second equality in the first line of Eq.~(\ref{Sigma0}). 
\section{Next-to-Leading Order Lagrangian}
\label{NLOLag}
With the discussion of the Goldstone manifold and its parameterization complete, the next step is to construct the explicit form of the Lagrangian for the multicomponent superfluid phase In order to do so, we set $\hat{\phi}_{1}=1$. This choice is arbitrary up to the constraint of $\hat{\phi}_{1}^{2}+\hat{\phi}_{2}^{2}+\hat{\phi}_{3}^{2}+\hat{\phi}_{4}^{2}=1$.
 
We also need an explicit form of the broken generators $X_{i}$, which we list below
\begin{equation}
\begin{split}
\label{generators}
&\textrm{Diquarks ($d_{\pm}$): }\\
&X_{1}=\frac{1}{\sqrt{2}}
\begin{pmatrix}
0&i\tau_{2}\\
-i\tau_{2}&0
\end{pmatrix},\ 
X_{2}=\frac{1}{\sqrt{2}}
\begin{pmatrix}
0&\tau_{2}\\
\tau_{2}&0
\end{pmatrix}\\
&\textrm{Charged Pions ($\pi_{\pm}$): }\\ 
&X_{3}=\frac{1}{\sqrt{2}}
\begin{pmatrix}
-\tau_{1}&0\\
0&-\tau_{1}
\end{pmatrix},\ 
X_{4}=\frac{1}{\sqrt{2}}
\begin{pmatrix}
-\tau_{2}&0\\
0&\tau_{2}
\end{pmatrix}\\
&\textrm{Neutral Pion ($\pi_{0}$):}\\
&X_{5}=\frac{1}{\sqrt{2}}
\begin{pmatrix}
\tau_{3}&0\\
0&\tau_{3}
\end{pmatrix},\
\end{split}
\end{equation}
with $X_{1,2}$ corresponding to broken generators for the diquarks, $X_{3,4}$ corresponding to the broken generators for the charged pions and $X_{5}$ corresponds to the broken generators of neutral pions respectively. We have chosen the following normalization
\begin{equation}
\begin{split}
\frac{1}{2}\Tr X_{i}X_{j}=\delta_{ij}\ ,
\end{split}
\end{equation}
similar to that of Gell-Mann matrices, the generators of $SU(3)$ flavor in $\chi$PT (or $SU(3)$ color in QCD). The broken generators, $X_{i}$, relevant to the multicomponent superfluid phase have been calculated using the broken generators of $SU(4)$ used in Ref.~\cite{Kogut:1999iv}, where the chiral limit is assumed, which means the broken generators are fully rotated from the normal chiral-symmetry-broken vacuum into a pure diquark phase with complete chiral restoration. We have (un)rotated the generators using $V_{\alpha}$ of Eq.~(\ref{Sigmaalpha}) with $\alpha=\pi/2$ to obtain the result in Eq.~(\ref{generators}).

Using these generators and Eq.~(\ref{Sigmaalpha}), we find the following contributions from the $\mathcal{O}(p^{n})$ Lagrangian, which we label $\mathcal{L}^{k}_{n}$, with $k$ indicating the mass dimension of the field fluctuations, $\phi_{i}$, for $i=1,2\dots 5$.
\begin{equation}
\begin{split}
\mathcal{L}_{2}^{0}&=f^{2}\left(2m_{j}^{2}\cos\alpha+\mu^{2}\sin^{2}\alpha\right)\ ,
\end{split}
\end{equation}
where
\begin{equation}
\begin{split}
\label{mj}
m_{j}^{2}=m^{2}+\frac{2Gj}{f^{2}}\tan\alpha\ .
\end{split}
\end{equation}
\begin{equation}
\begin{split}
\mathcal{L}_{2}^{1}&=-\sqrt{2}f\left (-m^{2}\sin\alpha+\mu^{2}\cos\alpha\sin\alpha\right.\\
&\left.+\frac{2Gj}{f^{2}}\cos\alpha\right )\phi_{1}-\sqrt{2}f\mu\sin\alpha\partial_{0}\phi_{2}\\
\end{split}
\end{equation}
\begin{equation}
\begin{split}
\label{L22}
\mathcal{L}_{2}^{2}&=\frac{1}{2}\partial_{\mu}\phi_{a}\partial^{\mu}\phi_{a}-\frac{1}{2}m_{a}^{2}\phi_{a}^{2}-\frac{i}{2}m_{12}\left (\phi_{1}\partial_{0}\phi_{2}-\phi_{2}\partial_{0}\phi_{1} \right )\\
&-\frac{i}{2}m_{34}\left (\phi_{3}\partial_{0}\phi_{4}-\phi_{4}\partial_{0}\phi_{3} \right )\\
\end{split}
\end{equation}
\begin{equation}
\begin{split}
\mathcal{L}_{2}^{3}&=\frac{1}{6\sqrt{2}f}\left [\left\{(4\mu^{2}\cos\alpha-m^{2})\sin\alpha+\frac{Gj}{f^{2}}\cos\alpha\right\}\right.\\
&\phi_{1}(\phi_{a}\phi_{a})
\left.-4\mu\phi_{1}\phi_{2}\partial_{0}\phi_{1}+4\mu(\phi_{3}^{2}\partial_{0}\phi_{2}+\phi_{4}^{2}\partial_{0}\phi_{2}\right.\\
&\left.+\phi_{5}^{2}\partial_{0}\phi_{2}-\phi_{2}\phi_{3}\partial_{0}\phi_{3}-\phi_{2}\phi_{4}\partial_{0}\phi_{4}-\phi_{2}\phi_{5}\partial_{0}\phi_{5})\right ]\\
\end{split}
\end{equation}
\begin{equation}
\begin{split}
\mathcal{L}_{2}^{4}&=\frac{1}{48f^{2}}(\phi_{a}\phi_{a})^{2}\left[\left (m_{j}^{2}\cos\alpha-4\mu^{2}\cos2\alpha\right )\phi_{1}^{2}\right.\\
&\left.+\left (m_{j}^{2}-4\mu^{2}\sin^{2}\alpha\right )(\phi_{2}^{2}+\phi_{3}^{2}+\phi_{4}^{2}+\phi_{5}^{2})\right]\\
&+\frac{1}{12f^{2}}\left [\phi_{a}\partial^{\mu}\phi_{a}\phi_{b}\partial^{\mu}\phi_{b}-\phi_{a}\phi_{a}\partial_{\mu}\phi_{b}\partial^{\mu}\phi_{b} \right]\\
&-\frac{1}{6f^{2}}\mu(\phi_{a}\phi_{a})\left [\cos\alpha(\phi_{1}\partial_{0}\phi_{2}-\phi_{2}\partial_{0}\phi_{1})\right.\\
&\left.+(\phi_{3}\partial_{0}\phi_{4}-\phi_{4}\partial_{0}\phi_{3}) \right ]\ .
\end{split}
\end{equation}
In $\mathcal{L}_{n}^{k}$, we have used the Einstein summation convention: all repeated indices ($a$ and $b$ in $\mathcal{L}_{2}^{2}$ and $\mathcal{L}_{2}^{4}$) are summed (from $1,2\dots 5$). We also note that the masses that appear in the quadratic term, $\mathcal{L}_{2}^{2}$, of Eq.~(\ref{L22}) are
\begin{equation}
\begin{split}
\label{masses}
m_{1}^{2}&=m_{j}^{2}\cos\alpha-\mu^{2}\cos2\alpha\\
m_{2}^{2}&=m_{j}^{2}\cos\alpha-\mu^{2}\cos^{2}\alpha\\
m_{12}&=2\mu\cos\alpha\\
m_{3}^{2}&=m_{j}^{2}\cos\alpha-\mu^{2}\cos^{2}\alpha\\
m_{4}^{2}&=m_{j}^{2}\cos\alpha-\mu^{2}\cos^{2}\alpha\\
m_{34}&=2\mu\\
m_{5}^{2}&=m_{j}^{2}\cos\alpha+\mu^{2}\sin^{2}\alpha\ ,
\end{split}
\end{equation}
where $\mu$ is the chemical potential, $f$ is the bare pion decay constant and
$m_{j}$ is defined in Eq.~(\ref{mj}), $m$ is the the bare diquark (or pion) mass and $j$ is the multicomponent superfluid source. We have replaced all the factors of $G\hat{m}/f^{2}$ that appear in $\mathcal{L}_{n}^{k}$ and $m_{a}$ through the scalar source term of Eq.~(\ref{sourceterm}) in $\mathcal{L}_{2}^{2}$ with the bare pion (diquark) mass $m$ using the zero-source ($j=0$) GOR relation of Eq.~(\ref{GOR}). It is also worth noting that in the presence of an external source $j$, the (bare) pion mass (squared) is modified as $m^{2}\cos\alpha\rightarrow m_{j}^{2}$ and $\alpha$ is non-zero due to the source term $j$ that rotates the vacuum into the multicomponent superfluid phase for all $\mu$.
Using the quadratic Lagrangian, it is straightforward to find the inverse propagator
\begin{equation}
\begin{split}
D^{-1}&=
\begin{pmatrix}
D^{-1}_{d}&0\\
0&D^{-1}_{\pi}
\end{pmatrix}\\
D_{d}^{-1}&=
\begin{pmatrix}
p^{2}-m_{1}^{2}&ip_{0}m_{12}\\
-ip_{0}m_{12}&p^{2}-m_{2}^{2}
\end{pmatrix}\\ 
D^{-1}_{\pi}&=
\begin{pmatrix}
p^{2}-m_{3}^{2}&ip_{0}m_{34}&0\\
-ip_{0}m_{34}&p^{2}-m_{4}^{2}&0\\
0&0&P^{2}-m_{5}^{2}
\end{pmatrix}\ ,
\end{split}
\end{equation}
where $D^{-1}$ is a $5\times 5$ matrix containing inverse propagators for diquarks and pions: $D_{d}^{-1}$ is the inverse propagator associated with diquark fluctuations $\phi_{1}$ and $\phi_{2}$; similarly, $D_{\pi}^{-1}$ is the inverse propagator associated with the pion fluctuations. $\phi_{3}$ and $\phi_{4}$ represent the fluctuations of the charged pion fields and $\phi_{5}$ represents the fluctuations of the neutral pions. 
$p^{2}$ is the 4-momentum squared in Minkowski space. We note that due to the presence of cross terms in the $1-2$ and $3-4$ components of the inverse propagators, only $m_{5}$ is the (bare) mass of the neutral pion. 
\subsection{Dispersion Relations}
\label{dispersionsection}
\begin{figure}[h]
    \centering
    \includegraphics[width=.5\textwidth]{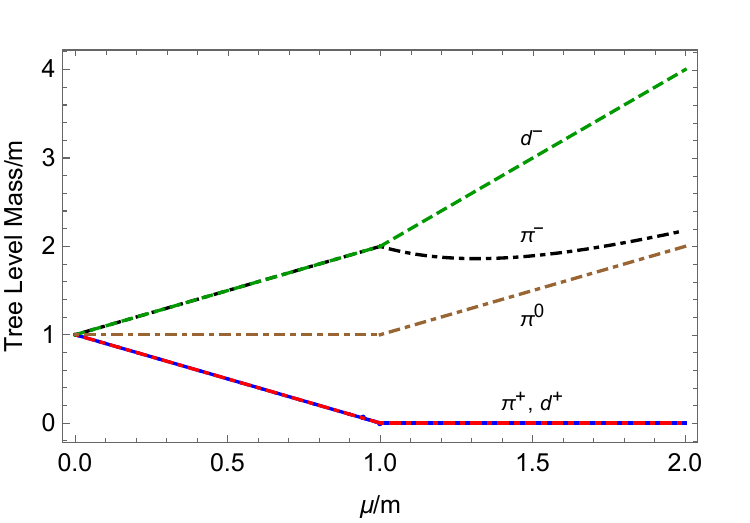}
    \caption{Tree-level masses of the five modes (pions and diquarks) when the diquark and isospin chemical potentials are equal, $\mu_{B}=\mu_{I}=\mu$. When $\mu>m$, both $\pi_{+}$ and $d_{+}$ become massless simultaneously. In the normal phase ($\mu<m$, the masses of $d_{-}$ and $\pi_{-}$ are degenerate and heavier than the masses of the other modes. The masses of $d_{+}$ and $\pi_{+}$ are also degenerate.}
\label{fig:spectrum}
\end{figure}

In order to find the remaining masses, we need the dispersion relations, which can be found by setting $\det D^{-1}=0$.
We get for the diquark and the pion dispersion relations (respectively)
\begin{equation}
\begin{split}
\label{dispersion}
E_{d_{\pm}}^{2}&=p^{2}+\frac{1}{2}(m_{1}^{2}+m_{2}^{2}+m_{12}^{2})\\
&\mp\frac{1}{2}\sqrt{4p^{2}m_{12}^{2}+(m_{1}^{2}+m_{2}^{2}+m_{12}^{2})^{2}-4m_{1}^{2}m_{2}^{2}}\\
E_{\pi_{\pm}}^{2}&=p^{2}+\frac{1}{2}(m_{3}^{2}+m_{4}^{2}+m_{34}^{2})\\
&\mp\frac{1}{2}\sqrt{4p^{2}m_{34}^{2}+(m_{3}^{2}+m_{4}^{2}+m_{34}^{2})^{2}-4m_{3}^{2}m_{4}^{2}}\\
&=p^{2}+m_{3}^{2}+\frac{1}{2}m_{34}^{2}\\
&\pm\frac{1}{2}\sqrt{4p^{2}m_{34}^{2}+(2m_{3}^{2}+m_{34}^{2})^{2}-4m_{3}^{4}}\\
E_{\pi_{0}}^{2}&=p^{2}+m_{5}^{2}\ ,
\end{split}
\end{equation}
where the masses $m_{i}$ ($i=1,2\dots 5$), $m_{12}$ and $m_{34}$ are defined in Eq.~(\ref{masses}). We find the tree-level masses of each of the modes by setting $p=0$ in the dispersion relation. We plot the masses (normalized by the bare diquark or pion mass) in Fig.~\ref{fig:spectrum}, noting that in the simultaneous condensed phase there are two massless modes $d_{+}$ and $\pi_{+}$ responsible for the multicomponent superfluidity observed when the diquark and isospin chemical potentials are equal. This is to be contrasted with the diquark and pion condensed phases where only one of the modes, either $d_{+}$ or $\pi_{+}$, becomes massless.
\\
We note that the positively charged diquark and the positively charged pion becomes massless for $\mu>m$ and are responsible for the multicomponent superfluid phase. While the diquark ($d_{+}$) carries a baryon charge, it does not carry an isospin charge; on the other hand the pion ($\pi_{+}$) carries an isospin charge but not a baryon charge. While the theory fixes the masses of each of the five particles, in the multicomponent superfluid, the condensate associated with only a non-zero baryon charge and the condensate associated with only a non-zero isospin charge are not fixed separately by the theory. This is evident in the analysis of Sec. \ref{review}: when $\mu_{B}=\mu_{I}$, the ground state of the system satisfies $\hat{\phi}_{1}^{2}+\hat{\phi}_{2}^{2}+\hat{\phi}_{3}^{2}+\hat{\phi}_{4}^{2}=1$, where subscripts $1,2$ label the diquarks and $3,4$ label the (electromagnetically) charged pions suggesting an expanded symmetry compared to the diquark (or pion) phase for which the constraint is consistent with a global $U(1)$ symmetry~\cite{Adhikari:2019mdk}.

In order to construct the one-loop effective potential, we will need to regulate the divergences in the one-loop diagrams, which requires counterterms generated through the tree-level $\mathcal{O}(p^{4})$ contribution to two-color, two-flavor $\chi$PT. The relevant $\chi$PT Lagrangian is~\footnote{We note that the counterterm proportional to $L_{0}$ can be written in terms of the counter-terms proportional to $L_{1}$, $L_{2}$ and $L_{3}$ in three-flavor, three-color $\chi$PT using the properties of the Gell-Mann matrices. This is not possible in two-color, two-flavor $\chi$PT~\cite{Splittorff:2001fy}.}
\begin{equation}
\begin{split}
\mathcal{L}_{4}&=L_{0}\Tr\left [\nabla_{\mu}\Sigma(\nabla_{\nu}\Sigma)^{\dagger}\nabla^{\mu}\Sigma(\nabla^{\nu}\Sigma)^{\dagger}\right ]\\
&+L_{1}\left (\Tr\left[\nabla_{\mu}\Sigma(\nabla^{\mu}\Sigma)^{\dagger}\right]\right )^{2}\\
&+L_{2}\Tr\left[\nabla_{\mu}\Sigma(\nabla_{\nu}\Sigma)^{\dagger}\right]\Tr\left[\nabla^{\mu}\Sigma(\nabla^{\nu}\Sigma)^{\dagger}\right]\\
&+L_{3}\Tr\left [\left (\nabla_{\mu}\Sigma(\nabla^{\mu}\Sigma)^{\dagger} \right )^{2} \right ]\\
&+L_{4}\Tr\left [\nabla_{\mu}\Sigma(\nabla^{\mu}\Sigma)^{\dagger} \right ]\Tr\left [\chi\Sigma^{\dagger}+\chi^{\dagger}\Sigma \right ]\\
&+L_{5}\Tr\left [\nabla_{\mu}\Sigma(\nabla^{\mu}\Sigma)^{\dagger}(\chi\Sigma^{\dagger}+\chi^{\dagger}\Sigma) \right ]\\
&+L_{6}\left [\Tr(\chi\Sigma^{\dagger}+\chi^{\dagger}\Sigma)\right ]^{2}+L_{7}\left [\Tr(\chi\Sigma^{\dagger}-\chi^{\dagger}\Sigma)\right ]^{2}\\
&+L_{8}\Tr(\Sigma\chi^{\dagger}\Sigma\chi^{\dagger}+\chi\Sigma^{\dagger}\chi\Sigma^{\dagger})+H_{2}\Tr(\chi\chi^{\dagger})\ ,
\end{split}
\end{equation}
where $L_{i}$ (with $i=1,2\cdots 8$) and $H_{2}$ are the low energy constants (LECs) of $\chi$PT that encode quark physics. The term proportional to $L_{7}$ is trivially zero in the two-color, two-flavor case. Using the parameterization of $\Sigma$ in Eq.~(\ref{Sigmaalpha}), we get the following static counterterms
\begin{equation}
\begin{split}
\label{L4static}
\mathcal{L}_{4}^{0}&=\left[4(L_{0}+L_{3})+16(L_{1}+L_{2}) \right]\mu^{4}\sin^{4}\alpha\\
&+8(4L_{4}+L_{5})\mu^{2}m_{j}^{2}\cos\alpha\sin^{2}\alpha\\
&+16(4L_{6}+L_{8})m^{4}_{j}\cos^{2}\alpha\\
&+4(H_{2}-2L_{8})(m_{j}^{4}+\bar{m}_{j}^{4})\ ,
\end{split}
\end{equation}
where we have defined as a new mass
\begin{equation}
\begin{split}
\bar{m}_{j}^{4}=4\left[(1-\tan^{2}\alpha)\left(\frac{Gj}{f^{2}}\right)^{2}-m^{2}\tan\alpha\frac{Gj}{f^{2}}\right]\ .
\end{split}
\end{equation}
$\mathcal{L}_{4}^{0}$ is the only counterterm required to renormalize the one-loop effective potential. However, for the sake of completeness, we also calculate the contributions with one and two field fluctuations ($\phi_{i}$) in \ref{L41L42}. These are required to renormalize the pion mass and the pion decay constant. Additionally, we need the explicit form of the running of the LECs which we define as
\begin{equation}
\begin{split}
L_{i}&=L_{i}^{r}(\Lambda)+\Gamma_{i}\lambda\\
H_{i}&=H_{i}^{r}(\Lambda)+\Delta_{i}\lambda\\
\lambda&=-\frac{\Lambda^{-2\epsilon}}{2(4\pi)^{2}}\left (\frac{1}{\epsilon}+1 \right )\\
\end{split}
\end{equation}
with $\Gamma_{0}=\frac{1}{48}$, $\Gamma_{1}=\frac{1}{32}$, $\Gamma_{2}=\frac{1}{16}$, $\Gamma_{3}=\frac{5}{48}$, $\Gamma_{4}=\frac{1}{16}$, $\Gamma_{5}=\frac{1}{4}$, $\Gamma_{6}=\frac{5}{128}$, $\Gamma_{7}=0$, $\Gamma_{8}=0$ and $\Delta_{2}=0$. Since $L_{i}$ is scale-independent, i.e.
\begin{equation}
\begin{split}
\lim_{\epsilon\rightarrow 0}\frac{dL_{i}}{d\Lambda}=0\ ,
\end{split}
\end{equation}
we can determine the running of $L_{i}^{r}$ with respect to the $\overline{MS}$ scale parameter $\Lambda$
\begin{equation}
\begin{split}
\frac{dL_{i}^{r}(\Lambda)}{d\Lambda}=-\frac{\Gamma_{i}}{(4\pi)^{2}\Lambda}\ .
\end{split}
\end{equation}
We can solve the differential equations above assuming $L_{i}^{r}$ is known at some reference scale $\Lambda_{0}$. This gives
\begin{equation}
\begin{split}
\label{LEC}
L_{i}^{r}(\Lambda)&=L_{i}^{r}(\Lambda_{0})-\frac{\Gamma_{i}}{2(4\pi)^{2}}\ln\frac{\Lambda^{2}}{\Lambda_{0}^{2}}\\
H_{i}^{r}(\Lambda)&=H_{i}^{r}(\Lambda_{0})-\frac{\Delta_{i}}{2(4\pi)^{2}}\ln\frac{\Lambda^{2}}{\Lambda_{0}^{2}}\ .
\end{split}
\end{equation}
Since $\Gamma_{7}=0$, $\Gamma_{8}=0$ and $\Delta_{2}=0$, the associated renormalized LECs ($L_{7}^{r}$, $L_{8}^{r}$ and $H_{2}^{r}$) are all scale-independent. In the rest of this work, we use the scale $\Lambda_{0}=m$ and use the notation $L_{i}^{r}(m)\equiv\bar{L}_{i}^{r}$ for convenience. 
\section{One-loop Effective Potential}
\label{1loopeffectivepotentialsection}
With all the ingredients in place, we can now calculate the one-loop effective potential of two-color, two-flavor $\chi$PT when the diquark chemical potential equals the isospin chemical potential, i.e. $\mu_{B}=\mu_{I}\equiv\mu$ using the Lagrangian and dispersion relations calculated in the previous section. For an analysis of the one-loop effective potential in the diquark phase, see Ref.~\cite{Kogut:2000ek}.
There are three contributions required to calculate the one-loop effective potential:
\begin{itemize}
\item Tree level contribution from the $\mathcal{O}(p^{2})$ $\chi$PT Lagrangian, 
\item Loop contribution from the $\mathcal{O}(p^{2})$ $\chi$PT Lagrangian, 
\item Tree level contribution from the $\mathcal{O}(p^{4})$ $\chi$PT Lagrangian.
\end{itemize}
The loop contributions from the $\mathcal{O}(p^{2})$ Lagrangian have UV divergences that need to be regulated. These divergences are cancelled exactly by the tree-level counterterms generated using the Lagrangian at $\mathcal{O}(p^{4})$ through $\mathcal{L}_{4}^{0}$ of Eq.~(\ref{L4static}). In order to compute the one-loop contribution from the $\mathcal{O}(p^{2})$ Lagrangian, we will use the $\overline{MS}$ scheme with dimensional regularization. The divergences that appear are in the ultraviolet ($UV$) but not in the infrared ($IR$) with the theory having both logarithmic and polynomial (up to quartic) divergences though only the former appear in dimensional regularization.

The $\mathcal{O}(p^{2})$ tree level contribution to the one-loop effective potential including the contribution of the source term $j$ is equal to $-\mathcal{L}_{2}^{0}$ of Eq.~(\ref{L4static}),
\begin{equation}
\begin{split}
\label{Vtree}
V_{\rm tree}^{j}&=f^{2}\left(2m_{j}^{2}\cos\alpha+\mu^{2}\sin^{2}\alpha\right)\ ,
\end{split}
\end{equation}
where $m_{j}^{2}$ is defined in Eq.~(\ref{mj}), and the one-loop contribution from the Lagrangian at $\mathcal{O}(p^{2})$ arises from the following divergent integral, which is a sum over the dispersion relations of Eq.~(\ref{dispersion})
\begin{equation}
\begin{split}
V_{\rm 1}^{j}&\equiv\frac{1}{2}\int_{p}\left (E_{d_{+}}+E_{d_{-}}+E_{\pi_{+}}+E_{\pi_{-}}+E_{\pi_{0}}\right )\\
\int_{p}&\equiv\left (\frac{e^{\gamma_{E}}\Lambda^{2}}{4\pi}\right )^{\epsilon}\int \frac{d^{d}p}{(2\pi)^{d}}\ ,
\end{split}
\end{equation}
where $\Lambda$ is the renormalization scale and $d=3-2\epsilon$. In order to isolate the divergences, we rewrite $V_{1}$ in terms of a divergence piece $V_{\rm div}$ and a finite piece $V_{\rm finite}$ as
\begin{equation}
\begin{split}
\label{Vfinite}
V_{1}^{j}=&V_{\rm div}^{j}+V_{\rm finite}^{j}\\
V_{\rm div}^{j}=&\frac{1}{2}\int_{p}\left (E_{1}+E_{2}+E_{3}+E_{4}+E_{\pi_{0}}  \right )\\
V_{\rm finite}^{j}=&\frac{1}{2}\int_{p}(E_{d_{+}}+E_{d_{-}}+E_{\pi_{+}}+E_{\pi_{-}})\\
-&\frac{1}{2}\int_{p}(E_{1}-E_{2}-E_{3}-E_{4})\\
\end{split}
\end{equation}
where $E_{i}=\sqrt{p^{2}+\overline{m}_{i}}$, with $i=1,2,3,4$ and $\overline{m}_{i}$ is defined as
\begin{equation}
\begin{split}
\overline{m}_{1}&\equiv\sqrt{m_{1}^{2}+\frac{1}{4}m_{12}^{2}}=\sqrt{m_{j}^{2}\cos\alpha+\mu^{2}\sin^{2}\alpha}=m_{5}\\
\overline{m}_{2}&\equiv\sqrt{m_{2}^{2}+\frac{1}{4}m_{12}^{2}}=\sqrt{m_{j}^{2}\cos\alpha}\\
\overline{m}_{3}&\equiv\sqrt{m_{3}^{2}+\frac{1}{4}m_{34}^{2}}=\sqrt{m_{j}^{2}\cos\alpha+\mu^{2}\sin^{2}\alpha}=m_{5}\\
\overline{m}_{4}&\equiv\sqrt{m_{4}^{2}+\frac{1}{4}m_{34}^{2}}=\sqrt{m_{j}^{2}\cos\alpha+\mu^{2}\sin^{2}\alpha}=m_{5}\\
\end{split}
\end{equation}
The choice of $E_{i}$ and $\overline{m}_{i}$ is motivated by the fact that the ultraviolet divergences in $E_{d_{+}}$, $E_{d_{-}}$, $E_{\pi_{+}}$ and $E_{\pi_{-}}$ are identical to the ultraviolet divergences in $E_{1}$, $E_{2}$, $E_{3}$ and $E_{4}$. This is easy to check by expanding these dispersion relations around $p=\infty$~\cite{Adhikari:2019mdk}. $V_{\rm div}$ is then calculated using the standard dimensional regularization integral $I_{1}$, of Eq.~(\ref{dimregint}). We get
\begin{equation}
\begin{split}
\label{Vdiv}
V_{\rm div}^{j}=&\frac{1}{2}\left [\sum_{i=1}^{4}I_{1}(\overline{m}_{i})+I_{1}(m_{5})\right ]\\
=&-\frac{1}{(4\pi)^{2}}\left [\frac{1}{\epsilon}+\frac{3}{2}+\log\left (\frac{\Lambda^{2}}{m_{5}^{2}} \right ) \right ]m_{5}^{4}\\
&-\frac{1}{4(4\pi)^{2}}\left [\frac{1}{\epsilon}+\frac{3}{2}+\log\left (\frac{\Lambda^{2}}{\bar{m}_{2}^{2}} \right ) \right ]\bar{m}_{2}^{4}\ ,
\end{split}
\end{equation}
where in the second line we have used the fact that $\overline{m}_{1}=\overline{m}_{3}=\overline{m}_{4}=m_{5}$. The ultraviolet divergences in the dispersion relations show up as poles in Eq. (\ref{Vdiv}), which need to be regulated. The static terms of the $\mathcal{O}(p^{4})$ Lagrangian that was calculated in Eq.~(\ref{L4static}) exactly cancel the divergences. The resulting loop contribution from the $\mathcal{O}(p^{2})$ Lagrangian and the tree-level contribution from the $\mathcal{O}(p^{2})$ Lagrangian is
\begin{equation}
\begin{split}
\label{V1}
V_{1}^{j}=&-\left[\frac{1}{(4\pi)^{2}}\left(\frac{1}{2}+\log\frac{\Lambda^{2}}{m_{5}^{2}}\right)+4(\bar{L}^{r}_{0}+\bar{L}^{r}_{2})+16(\bar{L}^{r}_{1}+\bar{L}^{r}_{3})\right]\mu^{4}\sin^{4}\alpha\\
&-\left[\frac{2}{(4\pi)^{2}}\left(\frac{1}{2}+\log\frac{\Lambda^{2}}{m_{5}^{2}}\right)+8(4\bar{L}^{r}_{4}+\bar{L}^{r}_{5})\right]\mu^{2}\cos\alpha\sin^{2}\alpha\\
&-\left[\frac{1}{(4\pi)^{2}}\left(\frac{5}{8}+\frac{1}{4}\log\frac{\Lambda^{2}}{\bar{m}_{2}^{2}}+\log\frac{\Lambda^{2}}{m_{5}^{2}}\right)+16(4\bar{L}^{r}_{6}+\bar{L}^{r}_{8})\right]m_{j}^{4}\cos^{2}\alpha\\
&-4(\bar{H}^{r}_{2}-2\bar{L}^{r}_{8})(m_{j}^{4}+\bar{m}_{j}^{4})\\
\end{split}
\end{equation}
where $\bar{L}_{i}^{r}=L_{i}^{r}(m)$ and $\overline{H}_{i}^{r}=H_{i}^{r}(m)$. The full one-loop effective potential is then
\begin{equation}
\begin{split}
\label{1loopeffectivepotential}
V^{j}_{\rm 1-loop}&=V^{j}_{\rm tree}+V^{j}_{\rm 1}+V^{j}_{\rm finite}\ ,
\end{split}
\end{equation}
where $V^{j}_{\rm tree}$, $V^{j}_{\rm 1}$ and $V^{j}_{\rm finite}$ can be found in Eqs.~(\ref{Vtree}), (\ref{Vfinite}) and (\ref{V1}).
\subsection{Ginzburg-Landau Expansion}
\noindent
As a non-trivial check of the one-loop effective potential, we consider the one-loop effective potential near the critical chemical potential. At tree-level, the critical chemical potential occurs at $\mu_{c}^{\rm tree}=m$, where $m$ is the pion mass at tree-level. The tree-level effective potential (with $j=0$) when expanded in powers of $\alpha$ gives
\begin{equation}
\begin{split}
V_{\rm tree}&=-2f^{2}m^{2}-f^{2}(\mu^{2}-m^{2})\alpha^{2}+\frac{1}{12}f^{2}(4\mu^{2}-m^{2})\alpha^{4}\ ,
\end{split}
\end{equation}
where we have ignored corrections of $\mathcal{O}(\alpha^{6})$. The result suggests a second order phase transition since the coefficient of the $\alpha^{4}$ term is positive (and equal to $m^{2}f^{2}/4$) at the tree level critical chemical potential. Quantum effects are known to not affect the order of the phase transition and the critical chemical potential as has been shown in two-flavor ($N_{c}=3$) $\chi$PT~\cite{Adhikari:2019mdk}, three-flavor ($N_{c}=3$) $\chi$PT~\cite{Adhikari:2018cea, Adhikari:2019mlf}, the quark-meson (QM) model~\cite{Adhikari:2018cea} and the NJL model~\cite{He:2005sp}. In order to explicitly verify this observation, we expand the one-loop effective potential in powers of $\alpha$ valid when the superfluid source $j\ll m$. We get the following Ginzburg-Landau (GL) effective potential~\cite{Adhikari:2019mdk}
\begin{equation}
\begin{split}
\label{V1loopGL}
V_{\rm 1-loop}^{j}(\alpha)&=c_{0}+c_{1}\alpha\phi+c_{2}\alpha^{2}+c_{4}\alpha^{4}+\mathcal{O}(\alpha^{6})\\
c_{0}&=-2m^{2}f^{2}\left[1+\frac{m^{2}}{16(4\pi f)^{2}}\right.\\
&\left.\left\{5+512\pi^{2}(\bar{H}^{r}_{2}+2(8\bar{L}^{r}_{6}+\bar{L}^{r}_{8})) \right\}\right]\\
c_{1}&=-2m^{2}f^{2}\left[1+\frac{16m^{2}}{f^{2}}\left(4\bar{L}^{r}_{6}+\bar{L}^{r}_{8}\right) \right]\\
c_{2}&=-2m^{2}f^{2}\left[\bar{\mu}\left\{1+\frac{8m^{2}}{f^{2}}(4L^{r}_{4}+L^{r}_{5})\right\}\right.\\
&\left.+\frac{4m^{2}}{f^{2}}\left\{4L^{r}_{4}+L^{r}_{5}-2(4L^{r}_{6}+L^{r}_{8}) \right\}\right]\\
c_{4}&=m^{2}f^{2}\left[ \frac{1}{4}+\frac{m^{2}}{16(4\pi f)^{2}}\left\{1\right.\right.\\
&\left.\left.-1024\pi^{2}\left \{\bar{L}^{r}_{0}+4(\bar{L}^{r}_{1}+\bar{L}^{r}_{2})+\bar{L}^{r}_{3}-4\bar{L}^{r}_{4}-\bar{L}^{r}_{5} \right \}\right\}\right ]\ ,
\end{split}
\end{equation}
where $\phi=\arctan(j/m)\approx j/m$ and $\bar{\mu}=(\mu-m)/\mu$. We can find the critical chemical potential for the phase transition to the multicomponent superfluid phase (at $j=0$) by setting the quadratic coefficient to zero, i.e. $c_{2}=0$. We get
\begin{equation}
\begin{split}
\mu_{c}^{\rm 1-loop}&=m\left [1+\frac{4m^{2}}{f^{2}}\left (-4\bar{L}_{4}^{r}-\bar{L}_{5}^{r}+8\bar{L}_{6}^{r}+2\bar{L}_{8}^{r}\right) \right ]\ ,
\end{split}
\end{equation}
which is exactly equal to the one-loop renormalized pion mass, $m_{\pi}$, calculated in \ref{selfenergy}, see Eq.~(\ref{renormalizedmass}). In order to verify that the phase transition is indeed second order at next-to-leading order, we need the coefficient of the $\alpha^{4}$ term evaluated at the physical pion mass, which equals $c_{4}$.
It  is greater than zero assuming the LECs are small, $\bar{L}_{i}^{r}\sim10^{-3}$, which is expected from three-color $\chi$PT and the $N_{c}$ power counting of the LECs~\cite{Bijnens:2014lea}. The coefficient, $c_{4}$, being positive guarantees that the critical chemical potential is a second order phase transition as is expected from tree-level calculations. Since $\chi$PT is an expansion in powers of $(4\pi f)^{-2}$, this should not be surprising. At tree-level, it is known that $c_{4}>0$ and $\mu_{I,c}^{\rm tree}=m$. The corrections due to loop effects are sub-leading in the chiral expansion and therefore not sufficient to change $c_{4}$ significantly from the tree level value of $\mathcal{O}(1)$ thus preserving the order of the phase transition.
A full analysis of the one-loop effective potential including regions with $\alpha\sim\mathcal{O}(1)$ requires the numerical values of the LECs in order to compute $V_{\rm finite}$ defined in Eq.~(\ref{Vfinite}). Even though $V_{\rm finite}$ cannot be integrated analytically, it is possible to find the ground state value of $\alpha$ at NLO near the critical isospin chemical potential ($\mu_{I}\gtrsim m_{\pi}$) where $\alpha\ll 1$. Minimizing the effective potential (with $j=0$) using the Ginzburg-Landau form of the one-loop effective potential calculated in Eq.~(\ref{V1loopGL}), we get
\begin{equation}
\begin{split}
\alpha_{\rm gs}&=\frac{16m^{2}}{f^{2}}\left(4\bar{L}^{r}_{4}+\bar{L}^{r}_{5}-8\bar{L}^{r}_{6}-2\bar{L}^{r}_{8}\right)\\
&+\bar{\mu}\left[4-\frac{m^{2}}{(4\pi f)^{2}}\left\{1-\frac{512\pi^{2}}{3}\left(6\bar{L}^{r}_{0}+24\bar{L}^{r}_{1}+24\bar{L}^{r}_{2}\right.\right.\right.\\
&\left.\left.\left.+6\bar{L}^{r}_{3}-28\bar{L}^{r}_{4}-7\bar{L}^{r}_{5}+32\bar{L}^{r}_{6}+8\bar{L}^{r}_{8}\right)\right\}\right ]\ ,
\end{split}
\end{equation}
where $\bar{\mu}=(\mu-m)/m$. Since the phase transition is second order with the critical chemical potential equal to the diquark (pion) mass, $\alpha_{gs}$ can be rewritten in the following form using Eq.~(\ref{renormalizedmass})
\begin{equation}
\begin{split}
\label{alphags}
\alpha_{\rm gs}&=\frac{4(\mu-m_{\pi})}{m_{\pi}}\left[1-\frac{m^{2}}{4(4\pi f)^{2}}\left\{1-\frac{256\pi^{2}}{3}\right.\right.\\
&\left.\left.(12\bar{L}^{r}_{0}+48\bar{L}^{r}_{1}+48\bar{L}^{r}_{2}+12\bar{L}^{r}_{3}-68\bar{L}^{r}_{4}-17\bar{L}^{r}_{5}+88\bar{L}^{r}_{6}+22\bar{L}^{r}_{8})\right\} \right ]\ .
\end{split}
\end{equation}
We note that $\alpha_{gs}$ is zero for $\mu\le m_{\pi}$ and becomes non-zero for $\mu>m_{\pi}$, which is consistent with a second order transition. 
In the following section, we can use $\alpha_{\rm gs}$ to calculate the chiral condensate, the multicomponent superfluid condensate and the number density near the critical isospin chemical potential, where the expression is valid since $\alpha\ll 1$.
\section{Condensates and Number Density}
\label{cond}
\noindent
Using the full one-loop effective potential of Eq.~(\ref{1loopeffectivepotential}), we can calculate formal expressions for the chiral condensate, the multicomponent superfluid condensate and the multicomponent superfluid density. Additionally, we also calculate the condensates and density near the phase transition using $\alpha_{\rm gs}$ and the GL one-loop effective potential of Eq.~(\ref{V1loopGL}).
\subsection{Chiral Condensate}
The chiral condensate can be calculated using the zero-source ($j=0$) one-loop effective potential and the GOR relation of Eq.~(\ref{GOR}). We get
\begin{equation}
\begin{split}
\label{ccnlo}
\langle \bar{\psi}\psi\rangle\equiv&\frac{\partial V^{j=0}_{\rm 1-loop}}{\partial \hat{m}}=\frac{G}{m f^{2}}\frac{\partial V^{j=0}_{\rm 1-loop}}{\partial m}\\
=&-4G\cos\alpha\\
&-\frac{4}{(4\pi f)^{2}}\left[1+\log\left (\frac{m^{2}}{m^{2}\cos\alpha+\mu^{2}\sin^{2}\alpha} \right )\right.\\
&\left.+64\pi^{2}(4\bar{L}^{r}_{4}+\bar{L}^{r}_{5})\right]\cos\alpha\sin^{2}\alpha G\mu^{2}\\
&-\frac{1}{2(4\pi f)^{2}}\left[5+2\log\sec\alpha\right.\\
&\left.+8\log\left(\frac{m^{2}}{m^{2}\cos\alpha+\mu^{2}\sin^{2}\alpha}\right)+8192\pi^{2}\bar{L}_{r}^{6}\right]\cos^{2}\alpha Gm^{2}\\
&-\frac{1}{2(4\pi f)^{2}}\left[512\pi^{2}(H^{r}_{2}+2\cos2\alpha \bar{L}^{r}_{8}\right]Gm^{2}\\
&-\frac{2}{(4\pi f)^{2}}\sin^{4}\alpha \frac{G\mu^{4}}{m^{2}}+\frac{G}{mf^{2}}\frac{\partial V^{j=0}_{\rm finite}}{\partial m}\ ,
\end{split}
\end{equation}
where the last term cannot be evaluated analytically. Near the critical chemical potential ($\mu_{I}\gtrsim m_{\pi}$), we use the GL one-loop effective potential of Eq.~(\ref{V1loopGL}) (since $V_{\rm finite}^{j=0}$ contributes) to calculate the chiral condensate analytically. We get
\begin{equation}
\begin{split}
&\lim_{\mu\rightarrow m_{\pi}}\langle \bar{\psi}\psi\rangle=\langle\bar{\psi}\psi\rangle_{0}\\
&-\left[\frac{1}{2}-\frac{4\mu^{2}}{f^{2}}\left (4\bar{L}_{4}^{r}+\bar{L}^{r}_{5} \right)+\frac{4m^{2}}{f^{2}}(4\bar{L}^{r}_{6}+\bar{L}^{r}_{8})\right]\alpha_{\rm gs}^{2}\ .
\end{split}
\end{equation}
where the leading corrections are of $\mathcal{O}(\alpha_{\rm gs}^{4})$ and $\langle\bar{\psi}\psi\rangle_{0}$ is the chiral condensate in the normal vacuum at one-loop:
\begin{equation}
\begin{split}
&\langle\bar{\psi}\psi\rangle_{0}=-4G\left [1+\frac{m^{2}}{32(4\pi f)^{2}}\left \{5+512\pi^{2}(\overline{H}_{2}^{r}+16\bar{L}^{r}_{6}+2\bar{L}^{r}_{8} \right \}\right]\ .
\end{split}
\end{equation}
\subsection{Multicomponent Superfluid Condensate}
\noindent
The multicomponent superfluid condensate is the derivative of $V_{\rm 1-loop}^{j}$ with respect to $j$ in the limit $j\rightarrow 0$, i.e.
\begin{equation}
\begin{split}
\label{sconlo}
\langle MSC \rangle&\equiv\left.\frac{\partial V^{j}_{\rm 1-loop}}{\partial j}\right |_{j=0}\\
&=-4G\sin\alpha\\
&-\frac{Gm^{2}}{(4\pi f)^{2}}\sin\alpha\cos\alpha\Big[-\log(\cos\alpha)\Big.\\
&\left.+4\log\left(\frac{m^{2}}{m^{2}\cos\alpha+\mu^{2}\sin^{2}\alpha} \right )+256\pi^{2}(4\bar{L}^{r}_{6}+\bar{L}^{r}_{8}) \right ]\\
&-\frac{4G\mu^{2}}{(4\pi f)^{2}}\sin^{3}\alpha\left [\log\left(\frac{m^{2}}{m^{2}\cos\alpha+\mu^{2}\sin^{2}\alpha}\right)\right.\\
&\left.+64\pi^{2}(\bar{L}^{r}_{4}+\bar{L}^{r}_{5}) \right ]+\left.\frac{\partial V_{\rm finite}^{j}}{\partial j}\right |_{j=0}\ ,
\end{split}
\end{equation}
where $V_{\rm 1-loop}^{j}$ is the effective one-loop potential in the presence of a multicomponent superfluid source and last term cannot be evaluated analytically. We note that at tree level the condensates satisfy the relation
\begin{equation}
\begin{split}
\label{circle}
\langle\bar{\psi}\psi\rangle_{\rm tree}^{2}+\langle MSC\rangle_{\rm tree}^{2}=(4G)^{2}\ ,
\end{split}
\end{equation}
where $-4G$ is the chiral condensate in the normal vacuum. As is clear from Eqs.~(\ref{ccnlo}) and (\ref{sconlo}) the condensates fail to satisfy the a similar constraint to Eq.~(\ref{circle}) at one-loop level.

Near the second order phase transition the multicomponent superfluid condensate is small and we can use the GL one-loop effective potential of Eq.~(\ref{V1loopGL}) to calculate it. We get
\begin{equation}
\begin{split}
\lim_{\mu_{I}\rightarrow m_{\pi}}\langle MSC\rangle=-4G\left[1+\frac{16m^{2}}{f^{2}}(4\bar{L}^{r}_{6}+\bar{L}^{r}_{8})\right]\alpha_{\rm gs}\ ,
\end{split}
\end{equation}
where $\alpha_{\rm gs}$ can be found in Eq.~(\ref{alphags}). The corrections are of $\mathcal{O}(\alpha_{\rm gs}^{2})$.
\subsection{Multicomponent Superfluid Number Density}
Finally we calculate the number density for the multicomponent superfluid.
\begin{equation}
\begin{split}
n\equiv&\frac{\partial V^{j=0}_{\rm 1-loop}}{\partial \mu}\\
=&-2f^{2}\mu\sin^{2}\alpha\\
&-\frac{1}{4\pi^{2}}\left [\log\left (\frac{m^{2}}{m^{2}\cos\alpha+\mu^{2}\sin^{2}\alpha} \right ) \right ]\mu m^{2}\sin^{2}\alpha\cos\alpha\\
&-\frac{1}{4\pi^{2}}\Big [\log\left (\frac{m^{2}}{m^{2}\cos\alpha+\mu^{2}\sin^{2}\alpha} \right )\Big.\\
&\Big.+64\pi^{2}(\bar{L}^{r}_{0}+4(\bar{L}^{r}_{1}+\bar{L}^{r}_{2}+\bar{L}^{r}_{3}) \Big ]\mu^{3}\sin^{4}\alpha\\
&+\frac{\partial V_{\rm finite}^{j=0}}{\partial \mu}\ ,
\end{split}
\end{equation}
where the last term cannot be evaluated analytically. Near the critical isospin chemical potential, we get
\begin{equation}
\begin{split}
\lim_{\mu\rightarrow m_{\pi}}n&=-2\mu\left [1+\frac{8m^{2}}{f^{2}} \right ]\alpha_{\rm gs}^{2}\ ,
\end{split}
\end{equation}
with corrections of $\mathcal{O}(\alpha_{\rm gs}^{4})$.
\section{Discussion and Future Work}
\label{discussion}
In this paper, we have studied two-color, two-flavor $\chi$PT (at next-to-leading order) when the diquark chemical potential is equal to the isospin chemical potentials. The theory is not accessible through lattice QCD due to the fermion sign problem -- even though the fermion determinant is real, it is not positive definite when both chemical potentials are present. We calculated the one-loop effective potential and used it to calculate the chiral condensate, multicomponent superfluid condensate and the multicomponent superfluid density. The original study of two-color, two-flavor $\chi$PT only characterized the diquark phase, which is a consequence of the breaking of the residual $U(1)$ symmetry when $\mu_{B}\neq\mu_{I}$~\cite{Kogut:1999iv}. When the diquark and isospin chemical potentials are equal the Lagrangian possesses has an enlarged $SU(2)$ symmetry, which is broken by the multicomponent superfluid phase. This phase is a ``mixture" of diquarks and pions with neither fixed contrary to the diquark or the pionic phases where the density of diquarks and pions are both fully determined by the theory. 

While we have limited our study to finding the full one-loop effective potential and then analyzing the second order phase transition, the chiral condensate, the multicomponent superfluid condensate and the number density near the phase transition in the loop expansion, the natural next step is to consider the theory away from the phase transition. Such an analysis requires the calculation of $V_{\rm finite}$ in Eq.~(\ref{Vfinite}), which depends on the bare diquark (pion) mass and the bare pion decay constant and the LECs of two-color, two-flavor $\chi$PT, which are yet to be determined. 

In future work, we will consider the possibility of determining the LECs of two-color, two-flavor $\chi$PT by fitting to lattice data at finite diquark density~\cite{Iida:2019rah,Braguta:2016cpw}. The fitting should be straightforward near the critical baryon chemical potential (with $\alpha\ll 1$) since the results are fully analytical with systematically known corrections but away from the critical chemical potential one has to consider the full NLO effects from $V_{\rm finite}$ of Eq.~(\ref{Vfinite}). The integral can only be performed numerically and depends on the ground state value of $\alpha_{\rm gs}$, which in turn depends on the LECs ($L_{i}^{r}$).
\section{Acknowledgement}
\noindent
P.A. would like to acknowledge Jens O. Andersen for many discussions on related work in Refs.~\cite{Adhikari:2019mlf, Adhikari:2019zaj, Adhikari:2019mdk}. H.N. would like to acknowledge research support provided through the Collaborative Undergraduate Research and Inquiry (CURI 2019) program at St. Olaf College.
\noindent
\appendix
\section{Useful Integrals in Dimensional Regularization}
\label{DR}
\noindent
We use the following notation
\begin{equation}
\begin{split}
\label{dimregint}
\int_{p}=\left ( \frac{e^{\gamma_{E}}\Lambda^{2}}{4\pi}\right )^{\epsilon}\int\frac{d^{d}p}{(2\pi)^{d}}\ ,
\end{split}
\end{equation}
to calculate some useful dimensional regularization integrals in $d=3-2\epsilon$ Euclidean dimensions,
\begin{equation}
\begin{split}
\label{dimregintegrals}
I_{1}(m^{2})&=\int_{p}\sqrt{p^{2}+m^{2}}\\
&=-\frac{m^{4}}{2(4\pi)^{2}}\left [\frac{1}{\epsilon}+\frac{3}{2}+\log\left (\frac{\Lambda^{2}}{m^{2}} \right ) \right ]\\
I_{a}(m^{2})&=\int_{p}\frac{1}{p^{2}-m^{2}}=-\frac{m^{2}}{(4\pi)^{2}}\left[\frac{1}{\epsilon}+1+\log\left (\frac{\Lambda^{2}}{m^{2}}\right ) \right]\\
I_{b}(m^{2})&=\int_{p}\frac{p^{2}}{p^{2}-m^{2}}=-\frac{m^{4}}{(4\pi)^{2}}\left[\frac{1}{\epsilon}+1+\log\left (\frac{\Lambda^{2}}{m^{2}}\right ) \right]\\
&=m^{2}I_{a}(m^{2})\ ,
\end{split}
\end{equation}
where the leading corrections are of $\mathcal{O}(\epsilon)$.
\section{$\mathcal{L}_{4}^{1}$ and $\mathcal{L}_{4}^{2}$}
\label{L41L42}
In this appendix, we list the $\mathcal{O}(p^{4})$ contributions to the Lagrangian of two-color, two-flavor $\chi$PT valid at finite density with $\mu=\mu_{B}=\mu_{I}$ with one ($k=1$) or two field ($k=2$) fluctuations. We adopt the notation $\mathcal{L}_{4}^{k}$ for $k=1,2$.
\begin{equation}
\begin{split}
\mathcal{L}_{4}^{1}&=\frac{8\sqrt{2}}{f}\left [2\left (4L_{6}+L_{8} \right)\cos\alpha m^{4}\right.\\
&\left.-\left(L_{4}+\frac{L_{5}}{4}\right)(1+3\cos2\alpha) m^{2}\mu^{2}\right.\\
&\left.-\left (L_{0}+4L_{1}+4L_{2}+L_{3} \right)\mu^{4}\sin^{2}\alpha\cos\alpha \right ]\sin\alpha\phi_{1}\\
&-\frac{8\sqrt{2}}{f}\mu\sin\alpha\Big [(4L_{4}+L_{5})\cos\alpha\Big.\\
&\left.+(L_{0}+4L_{1}+4L_{2}+L_{3})\mu^{2}\sin^{2}\alpha\right.\\
&\left.+\frac{Gj}{f^{2}}2(4L_{4}+L_{5})\sin\alpha\right ]\partial_{0}\phi_{2}\\
\end{split}
\end{equation}
\begin{equation}
\begin{split}
\mathcal{L}_{4}^{2}&=-\frac{8(4L_{6}+L_{8})}{f^{2}}\left [\cos2\alpha\phi_{1}^{2}+\cos^{2}\alpha(\phi_{a}\phi_{a}-\phi_{1}^{2}) \right ]m^{4}\\
&+\frac{4L_{4}+L_{5}}{2f^{2}}\left [(-\cos\alpha+9\cos3\alpha)\phi_{1}^{2}\right.\\
&\left.+(5\cos\alpha+3\cos3\alpha)(\phi_{2}^{2}+\phi_{3}^{2}+\phi_{4}^{2})\right.\\
&\left.-(12\sin^{2}\alpha\cos\alpha)\phi_{5}^{2} \right ]m^{2}\mu^{2}\\
&+\frac{4(L_{0}+L_{3})+16(L_{1}+L_{2})}{f^{2}}\left [(1+2\cos2\alpha)\sin^{2}\alpha\phi_{1}^{2}\right.\\
&\left.+\sin^{2}\alpha\cos^{2}\alpha(\phi_{2}^{2}+\phi_{3}^{2}+\phi_{4}^{2})-\sin^{4}\alpha\phi_{5}^{2}\right ]\mu^{4}\\
&+\frac{32(4L_{6}+L_{8})}{f^{2}}\left [\cos2\alpha\phi_{1}^{2}-\sin^{2}\alpha(\phi_{a}\phi_{a}-\phi_{1}^{2}) \right ]\left(\frac{Gj}{f^{2}}\right)^{2}\\
&-\frac{16(4L_{6}+L_{8})}{f^{2}}\left [2\sin2\alpha\phi_{1}^{2}+\sin2\alpha(\phi_{a}\phi_{a}-\phi_{1}^{2}) \right ]\left (\frac{Gj}{f^{2}} \right )m^{2}\\
&-\frac{4L_{4}+L_{5}}{f^{2}}\left[3(\sin\alpha-3\sin3\alpha)\phi_{1}^{2}\right.\\
&\left.+(\sin\alpha-3\sin3\alpha)(\phi_{2}^2+\phi_{3}^2+\phi_{4}^2)+12\sin^{3}\alpha\phi_{5}^{2}\right]\left(\frac{Gj}{f^{2}}\right)\mu^{2}\\
&+\frac{4}{f^{2}}\bigg [(4L_{4}+L_{5})m^{2}\cos\alpha+\left\{L_{0}+L_{3}+4(L_{1}+L_{2})\right\}\mu^{2}\sin^{2}\alpha\bigg.\\
&\left.+2(4L_{4}+L_{5})\sin\alpha\frac{Gj}{f^{2}} \right ]\partial_{\mu}\phi_{a}\partial^{\mu}\phi_{a}\\
&+\frac{8}{f^{2}}\bigg[\left \{L_{0}+L_{3}+4(L_{1}+L_{2})\right\}\mu^{2}\sin^{2}\alpha\bigg]\partial_{\mu}\phi_{2}\partial^{\mu}\phi_{2}\\
&+\frac{8}{f^{2}}\bigg [3\left\{L_{0}+L_{3}+4(L_{1}+L_{2})\right\}\mu^{3}\sin^{2}\alpha\cos\alpha\bigg.\\
&\left.+(4L_{4}+L_{5})\mu m^{2}\cos2\alpha+2(4L_{4}+L_{5})\frac{Gj}{f^{2}}\sin2\alpha \right ]\phi_{1}\partial_{0}\phi_{2}\\
&-\frac{8}{f^{2}}\bigg [\left\{L_{0}+L_{3}+4(L_{1}+L_{2})\right\}\mu^{3}\sin^{2}\alpha\cos\alpha\bigg.\\
&\bigg.+(4L_{4}+L_{5})\mu m^{2}\cos^2\alpha+(4L_{4}+L_{5})\frac{Gj}{f^{2}}\sin2\alpha \bigg ]\phi_{2}\partial_{0}\phi_{1}\\
&+\frac{8}{f^{2}}\bigg [\left\{L_{0}+L_{3}+4(L_{1}+L_{2}) \right\}\mu^{3}\sin^{2}\alpha\bigg.\\
&\left.+(4L_{4}+L_{5})\mu m^{2}\cos\alpha+2(4L_{4}+L_{5})\frac{Gj}{f^{2}}\mu\sin\alpha \right ]\\
&(\phi_{3}\partial_{0}\phi_{4}-\phi_{4}\partial_{0}\phi_{3})\ .
\end{split}
\end{equation}
$\mathcal{L}^{1}_{4}$ with $\alpha=0$ is required for the calculation of the pion self-energy and the renormalized pion decay constant in \ref{selfenergy}.
\section{One-loop pion mass and pion decay constant}
\label{selfenergy}
\noindent
In order to find the pion self-energies in the normal vacuum, we need the following contributions from the $\mathcal{O}(p^{2})$ and $\mathcal{O}(p^{4})$ Lagrangian, 
\begin{equation}
\begin{split}
\mathcal{L}_{2}^{4}&=\frac{m^{2}}{48f^{2}}(\phi_{a}\phi_{a})^{2}\\
&+\frac{1}{12f^{2}}\left [\phi_{a}\partial^{\mu}\phi_{a}\phi_{b}\partial^{\mu}\phi_{b}-\phi_{a}\phi_{a}\partial_{\mu}\phi_{b}\partial^{\mu}\phi_{b} \right]\\
\mathcal{L}_{4}^{2}&=\frac{(16L_{4}+4L_{5})m^{2}}{f^{2}}\left (\partial_{\mu}\phi_{a}\partial^{\mu}\phi_{a} \right )\\
&-\frac{(32L_{6}+8L_{8})m^{4}}{f^{2}}\phi_{a}\phi_{a}
\end{split}
\end{equation}
The self-energy for $\phi_{1}$ is
\begin{equation}
\begin{split}
-i\Sigma(p^{2})&=i\left [\frac{m^{2}}{48f^{2}}\left \{12I_{a}(m^{2})+2\times 2\times 4I_{a}(m^{2}) \right\}\right.\\
&\left.-\frac{1}{12f^{2}}\left\{p^{2}2\times 4I_{a}(m^{2})+2\times4 I_{b}(m^{2})\right\}\right ]\ ,
\end{split}
\end{equation}
where $I_{a}$ and $I_{b}$ are defined in Eq.~(\ref{dimregintegrals}). The self-energy counter-term for $\phi_{1}$ is
\begin{equation}
\begin{split}
-i\Sigma_{\rm ct}(p^{2})&=i\left[2\times \frac{(16L_{4}+4L_{5})m^{2}p^{2}}{f^{2}}\right.\\
&\left.-2\times \frac{(32L_{6}+8L_{8})m^{4}}{f^{2}}\right]\ ,
\end{split}
\end{equation}
where $f$ is the bare pion decay constant, $m$ is the bare pion mass and $\bar{L}^{r}_{i}$ are the LECs defined in Eq.~(\ref{LEC}).
We note that the divergences in $\Sigma$ are canceled exactly by those in the self-energy counter-term $\Sigma_{\rm ct}$
\begin{equation}
\begin{split}
\label{renormalizedmass}
m_{\pi}^{2}&=m^{2}+\Sigma(m^{2})+\Sigma_{\rm ct}(m^{2})\\
&=m^{2}\left [1-\frac{8m^{2}}{f^{2}}(4\bar{L}^{r}_{4}+\bar{L}^{r}_{5}-2(4\bar{L}_{6}^{r}+\bar{L}_{8}^{r})) \right ]\ .
\end{split}
\end{equation}
The pion decay constant can be calculated by considering the matrix that couples the left current with a single pion field~\cite{scherer2011primer}. We get
\begin{equation}
\begin{split}
f_{\pi}^{2}&=f^{2}\left [1+\frac{8m^{2}}{f^{2}}(4\bar{L}_{4}^{r}+\bar{L}_{5}^{r})\right ]\ ,
\end{split}
\end{equation}
where $f$ is the bare pion decay constant, $m$ is the bare pion mass and $\bar{L}^{r}_{i}$ are the LECs defined in Eq.~(\ref{LEC}).
\bibliographystyle{apsrev4-1}
\bibliography{/Users/prabal7e7/Documents/Research/bib.bib}
\end{document}